\documentclass[]{spie}  

\usepackage{amsmath,amsfonts,amssymb}
\usepackage{graphicx}
\usepackage{placeins}
\usepackage[colorlinks=true, allcolors=blue]{hyperref}

\title{2022 Upgrade and Improved Low Frequency Camera Sensitivity for CMB Observation at the South Pole}

\author[a]{A. Soliman}
\author[b]{P.A.R. Ade}
\author[c]{Z. Ahmed}
\author[d]{M. Amiri}
\author[e]{D. Barkats}
\author[f]{R. Basu Thakur}
\author[g]{C.A. Bischoff}
\author[c, h]{D. Beck}
\author[f, i]{J.J. Bock}
\author[j]{V. Buza}
\author[k]{J. Cheshire}
\author[q, l]{J. Connors}
\author[e]{J. Cornelison}
\author[m]{M. Crumrine}
\author[h, c]{A.J. Cukierman}
\author[l]{E.V. Denison}
\author[e]{M.I. Dierickx}
\author[n]{L. Duband}
\author[e]{M. Eiben}
\author[d]{S. Fatigoni}
\author[o, p]{J.P. Filippini}
\author[g]{C. Giannakopoulos}
\author[h]{N. Goeckner-Wald}
\author[e]{D.C. Goldfinger}
\author[h]{J. Grayson}
\author[e]{P.K. Grimes}
\author[m]{G. Hall}
\author[h]{G. Halal}
\author[d]{M. Halpern}
\author[g]{E. Hand}
\author[e]{S.A. Harrison}
\author[c]{S. Henderson}
\author[f, i]{S.R. Hildebrandt}
\author[l]{G.C. Hilton}
\author[l]{J. Hubmayr}
\author[f]{H. Hui}
\author[h, c]{K.D. Irwin}
\author[h, f]{J. Kang}
\author[e, j]{K.S. Karkare}
\author[f]{S. Kefeli}
\author[e, q]{J.M. Kovac}
\author[h, c]{C.L. Kuo}
\author[m]{K. Lau}
\author[j]{E.M. Leitch}
\author[o]{A. Lennox}
\author[h]{T. Liu}
\author[i]{K.G. Megerian}
\author[f]{L. Minutolo}
\author[f]{L. Moncelsi}
\author[h, k]{Y. Nakato}
\author[r]{T. Namikawa}
\author[i]{H.T. Nguyen}
\author[f, i]{R. O'Brient}
\author[g]{S. Palladino}
\author[e]{M.A. Petroff}
\author[k]{N. Precup}
\author[n]{T. Prouve}
\author[m, k]{C. Pryke}
\author[e ,s]{B. Racine}
\author[l]{C.D. Reintsema}
\author[h]{M. Salatino}
\author[f]{A. Schillaci}
\author[e]{B.L. Schmitt}
\author[k]{B. Singari}
\author[e, q]{T. St. Germaine}
\author[f]{B. Steinbach}
\author[b]{R.V. Sudiwala}
\author[h, c]{K.L. Thompson}
\author[e]{C. Tsai}
\author[b]{C. Tucker}
\author[i]{A.D. Turner}
\author[g, o]{C. Umilt\`{a}}
\author[e]{C. Verg\`{e}s}
\author[t, j]{A.G. Vieregg}
\author[f]{A. Wandui}
\author[i]{A.C. Weber}
\author[d]{D.V. Wiebe}
\author[m]{J. Willmert}
\author[c]{W.L.K. Wu}
\author[h]{H. Yang}
\author[h, c]{K.W. Yoon}
\author[h, c]{E. Young}
\author[h]{C. Yu}
\author[e]{L. Zeng}
\author[f]{C. Zhang}
\author[f]{S. Zhang}

\affil[a]{Department of Engineering and Applied Science, California Institute of Technology, Pasadena, CA 91125, USA}

\affil[b]{School of Physics and Astronomy, Cardiff University, Cardiff, CF24 3AA, United Kingdom}
\affil[c]{Kavli Institute for Particle Astrophysics and Cosmology, SLAC National Accelerator Laboratory, Menlo Park, California 94025, USA}
\affil[d]{Department of Physics and Astronomy, University of British Columbia, Vancouver, British Columbia V6T 1Z1, Canada}
\affil[e]{Center for Astrophysics, Harvard \& Smithsonian, Cambridge, Massachusetts 02138, U.S.A}
\affil[f]{Department of Physics, California Institute of Technology, Pasadena, California 91125, USA}
\affil[g]{Department of Physics, University of Cincinnati, Cincinnati, Ohio 45221, USA}
\affil[h]{Department of Physics, Stanford University, Stanford, California 94305, USA}
\affil[i]{Jet Propulsion Laboratory, Pasadena, California 91109, USA}
\affil[j]{Kavli Institute for Cosmological Physics, University of Chicago, Chicago, Illinois 60637, USA}
\affil[k]{Minnesota Institute for Astrophysics, University of Minnesota, Minneapolis, Minnesota, 55455, USA}
\affil[l]{National Institute of Standards and Technology, Boulder, Colorado 80305, USA}
\affil[m]{School of Physics and Astronomy, University of Minnesota, Minneapolis, Minnesota 55455, USA}
\affil[n]{Service des Basses Temp\'{e}ratures, Commissariat \`{a} l'Energie Atomique, 38054 Grenoble, France}
\affil[o]{Department of Physics, University of Illinois at Urbana-Champaign, Urbana, Illinois 61801, USA}
\affil[p]{Department of Astronomy, University of Illinois at Urbana-Champaign, Urbana, Illinois 61801, USA}
\affil[q]{Department of Physics, Harvard University, Cambridge, Massachusetts 02138, USA}
\affil[r]{Kavli Institute for the Physics and Mathematics of the Universe (WPI), UTIAS, The University of Tokyo, Kashiwa, Chiba 277-8583, Japan}
\affil[s]{Aix-Marseille Universit\'{e}, CNRS/IN2P3, CPPM, 13288 Marseille, France}
\affil[t]{Department of Physics, Enrico Fermi Institute, University of Chicago, Chicago, Illinois 60637, USA}

\authorinfo{Corresponding author: A. Soliman, asoliman@caltech.edu}

\begin{document} 
\maketitle

\begin{abstract}
Constraining the Galactic foregrounds with multi-frequency Cosmic Microwave Background (CMB) observations is an essential step towards ultimately reaching the sensitivity to measure primordial gravitational waves (PGWs), the sign of inflation after the Big-Bang that would be imprinted on the CMB. The BICEP Array is a set of multi-frequency cameras designed to constrain the energy scale of inflation through CMB B-mode searches while also controlling the polarized galactic foregrounds. The lowest frequency BICEP Array receiver (BA1) has been observing from the South Pole since 2020 and provides 30 GHz and 40 GHz data to characterize galactic synchrotron in our CMB maps.  In this paper, we present the design of the BA1 detectors and the full optical characterization of the camera including the on-sky performance at the South Pole. The paper also introduces the design challenges during the first observing season including the effect of out-of-band photons on detectors performance. It also describes the tests done to diagnose that effect and the new upgrade to minimize these photons, as well as installing more dichroic detectors during the 2022 deployment season to improve the BA1 sensitivity. We finally report background noise measurements of the detectors with the goal of having photon-noise dominated detectors in both optical channels. BA1 achieves an improvement in mapping speed compared to the previous deployment season. 

\end{abstract}


\section{INTRODUCTION}
\label{sec:intro}  

The primordial gravitational waves (PGWs) have been predicted to cause a  unique polarization signature in the Cosmic Microwave Background (CMB). The ultimate detection of these PGWs could open a new window on physics of the newborn universe since their detection would be direct evidence for the theory of cosmic inflation. To constrain the PGWs magnitude, and thus the energy scale of inflation, we must account for polarized Galactic foregrounds \cite{sync1} with multi-frequency maps of the CMB. To date, BICEP/Keck (BK) program has placed the strongest constrains on the amplitude of PGWs, the tensor-to-scalar ratio \emph{r} \cite{BKI}. The current sensitivity is limited by the modeling of both gravitational lensing and polarized Galactic foregrounds. The BICEP/Keck team has collaborated with the SPT-3G team to develop the delensing techniques \cite{delensing} in the CMB maps  to improve the constraints on r. The Galactic foregrounds characterization, especially synchrotron emission remains a challenge beyond that collaboration and we need a high sensitive telescope at low frequency channels (30/40 GHz) since it dominates at these frequency bands.  

BICEP Array telescope \cite{hui18} has been developed with an exceptional sensitivity and wide frequency coverage to look for inflationary signals. It represents the latest advanced telescope to map the polarization of CMB  over 30/40, 95, 150, and 220/270 GHz channels to fully characterize the Galactic synchrotron and thermal dust emissions. Thermal Kinetic Inductance Detectors (TKIDs) \cite{albert} are currently being developed and tested to be used for higher frequencies BICEP Array receivers.

The first BICEP Array receiver (BA1) has been deployed to the Amundsen-Scott South Pole Station during the 2020 austral season and is currently measuring the polarized Galactic synchrotron radiation at 30/40 GHz. We use the CMB maps at 30/40 GHz to constrain the synchrotron foreground components in the most sensitive 95/150 GHz bands closer to the CMB peak frequency which will enhance the sensitivity on r. 

We have tested BA1 during 2020 and 2021 seasons of CMB observations. In this paper, we report the performance update of the BA1 camera and design challenges associated with these deployment seasons. We will also discuss our recent upgrades during that 2022 deployment season to improve the sensitivity of  BA1 camera and overall receiver. We upgraded the focal plane with more dichroic detector tiles for higher detector counts and an improved filter configuration to eliminate the high frequency leaking power. These upgrades will help us to improve the mapping speed and better characterize the synchrotron contamination to CMB B-modes at the level requested to potentially measure the primordial gravitational waves.

\section{Design Overview and Optical Performance}

\subsection{Detector Design }
\label{sec:title}

\begin{figure} [ht]
   \begin{center}
   \begin{tabular}{c} 
\includegraphics[height=10cm]{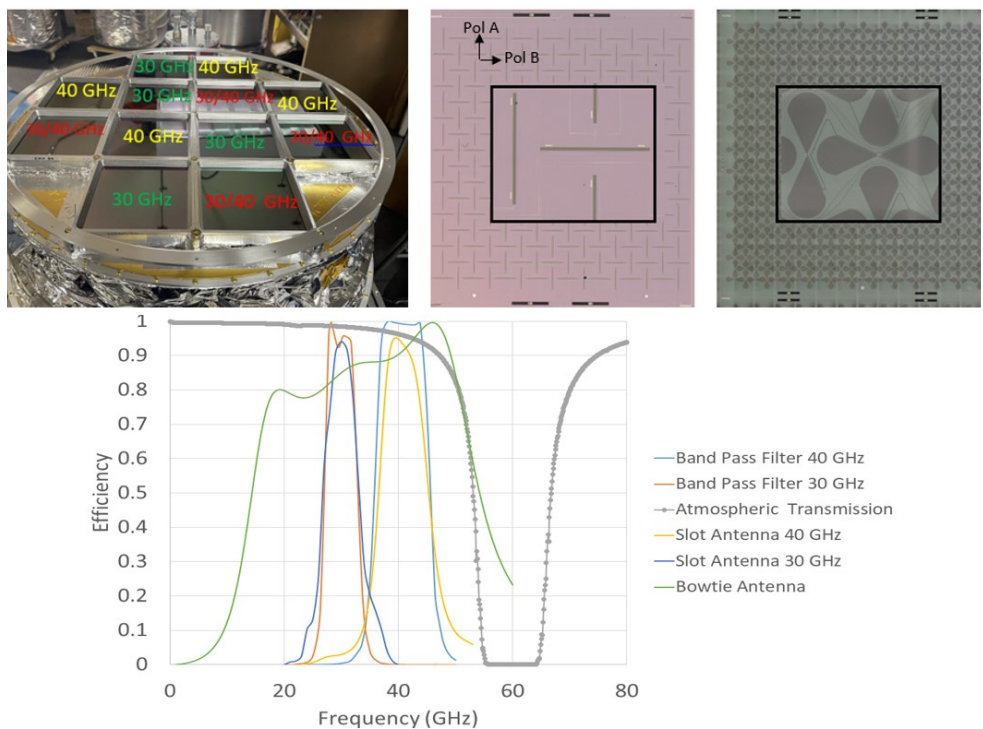}
   \end{tabular}
   \end{center}
   \caption[example] 
   { \label{fig:first} Top: The BA1 focal plane camera configuration during 2022 season houses
   a mixture of single band slot detector modules and dual color Bowtie detector modules for higher detector counts.  Bottom: Bands of antennas and filters as individually simulated in HFSS and Sonnet. We co-plot these with the atmospheric transmission to show that these do not overlap the CO2 emissions. }
   \end{figure}

Antenna-coupled transition-edge sensor (TES) detectors have been used in numerous successful CMB experiments, including BICEP Array. \cite{Ale20}\cite{Soliman20}\cite{Cheng20} Our pixel design  contains two orthogonal polarized detectors, each consisting of antenna array coherently combined through a summing network and connected to a superconducting transition edge sensor (TES) bolometers \cite{Cheng20}. Furthermore, we use on-chip band defining filters designed to select the frequency of interest and reject out-of-band signals, specifically those on atmospheric lines. We use arrays of rectangular slot antennas\cite{BKII} for single color detectors and arrays of broadband bowties for dual color pixels. In the dual color pixels, the filters are part of a diplexer circuit that partitions the antenna’s broad bandwidth into narrow photometric channels at 30 GHz and 40 GHz. A novel broadband corrugated frame\cite{soli18} has been used around the wafer edges to minimize the temperature-to-polarization leakage in the CMB maps. The HFSS simulation of the antenna array and sonnet simulation of the band-pass filter show that the detectors have been designed for frequency band centered at 30 GHz and 40 GHz with about 26\% fractional bandwidth (Fig. \ref{fig:first}). 

\label{sec:title}
   
\FloatBarrier
\subsection{Full Optical Characterization}
\label{sec:title}
\begin{figure} [ht]
   \begin{center}
   \begin{tabular}{c} 
   \includegraphics[height=5cm]{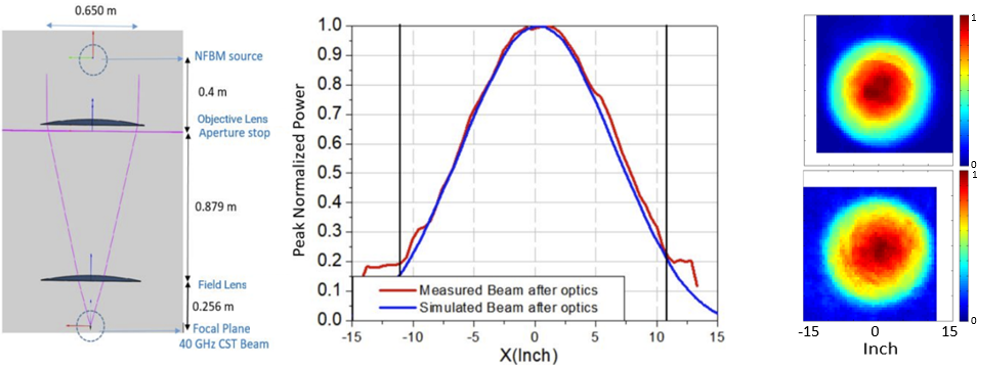}
   \end{tabular}
   \end{center}
   \caption[example] 
   { \label{fig:second} Left: The simple optics model of BA1 in Grasp Software. Middle: The resulting output beam planner cut of Grasp simulation shows a good agreement with the measured beam cut of 40 GHz detector. Right: Per-detector composite near-field beam maps of the telescope, obtained by co-adding all individual NFBMs for a detector module, 40 GHz detector (top) and 30 GHz detector (bottom). The beams are peak normalized.}
   \end{figure}

We measure the detector beams both in the lab and the field to verify our instrument performance and to inform our analysis.
We measure the near-field beam maps (NFBMs) with a chopped source on a pair of translation stages in front of the BA1 camera aperture plane and record the detector timestreams. The NFBMs help us to verify performance and monitor any potential issues with the on-chip millimeter wave circuits. We have demonstrated the good agreement between the measured and simulated beams on the near-field of the detectors without the optical elements \cite{Soliman20}\cite{soli18} . Additionally, we simulated our antenna’s beam with CST MICROWAVE STUDIO software and then used the GRASP software package to propagate that beam through the BA1 optics\cite{hui18}. The resulting simulated beam agrees well with the measured beams (Fig. \ref{fig:second}). The black vertical lines indicate location of the aperture stop. We noticed that the measured beam (red curve) doesn't drop beyond the aperture stop and we thought that the reflection around the aperture stop may have caused that issue. The measured optically-active beam maps per a detector module have been coadded together to create the per-detector composite beam map at each frequency band as shown in the right side of Fig. \ref{fig:second}.

\begin{figure} [ht]
   \begin{center}
   \begin{tabular}{c} 
   \includegraphics[height=5cm]{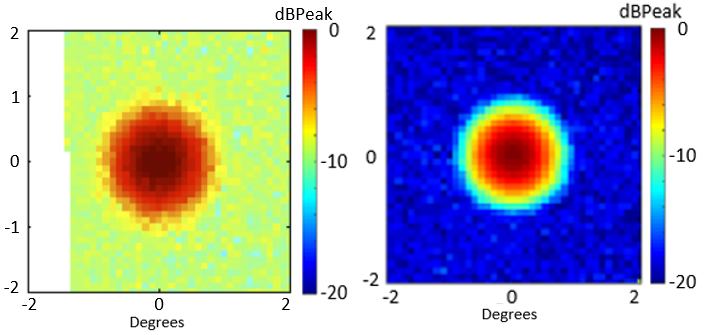}
   \end{tabular}
   \end{center}
   \caption[example] 
   { \label{fig:third} The measured far-field beam maps on the sky at the South Pole. Left: 30GHz detector. Right: 40GHz detector }
   \end{figure}

We also measured the far-field beams of BA1 camera on the sky at the South Pole (Fig. \ref{fig:third}) by using a thermal chopped source about 215 m away and raster the telescoped beam over the source with the telescope drive motors\cite{beam1}. The measured Gaussian beamwidth of 40 GHz and 30 GHz detectors are 0.36$^{\circ}$ and 0.47$^{\circ}$, respectively which are consistent with the scaled BICEP3 beamwidth at 95 GHz \cite{beam} with less than 10\% error. The simulations and measurements validate the detector and optical design of BA1 camera.  

\section{IMPROVED Camera SENSITIVITY (2020 vs 2022 data)}
\subsection{Out of Band Dark Loading}
\begin{figure} [ht]
   \begin{center}
   \begin{tabular}{c} 
   \includegraphics[height=5.5cm]{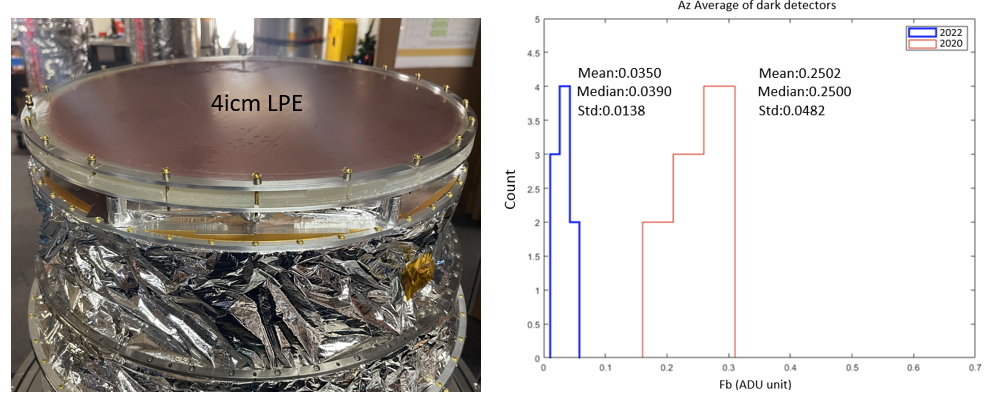}
   \end{tabular}
   \end{center}
   \caption[example] 
   { \label{fig:fourth} Left: The BA1 focal plane with the 4icm LPE filter upgrade during 2022 season. Right: Response of dark detectors to a chopped source for the 2020 deployment season versus 2022 deployment season. The only difference is an addition of a 4icm LPE filter that  highly minimizes the Blue leak.}
   \end{figure}

\begin{figure} [ht]
   \begin{center}
   \begin{tabular}{c} 
   \includegraphics[height=9.5cm]{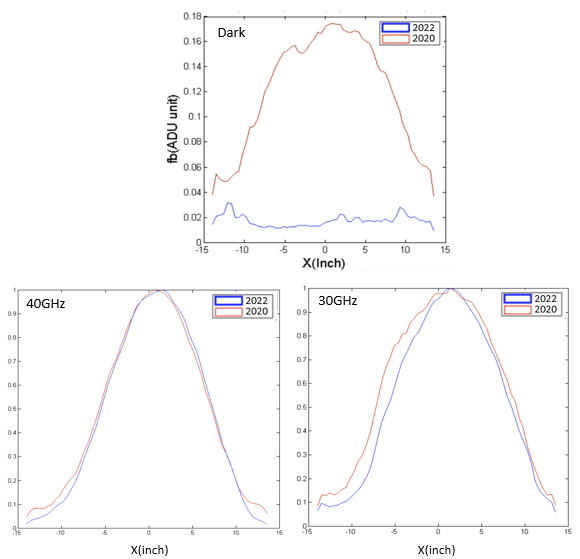}
   \end{tabular}
   \end{center}
   \caption[example] 
   { \label{fig:fifth}  The beam profiles of 2020 season versus 2022 season after the 4icm LPE upgrade. Top: Dark Detector (the additional filter largely eliminates the direct stimulation). Bottom/Left: 40 GHz detector. Bottom/Right: 30 GHz detector.  
}
   \end{figure}

BICEP detector arrays include a small number of dark detectors where the TES bolometers are not connected to an antenna.  We use these dark detectors for noise and loading studies. During the 2020 season, the dark detectors in our focal plane experienced excess loading that we understood to be direct stimulation of the bolometers with high frequency out-of-band power (a Blue leak). We performed  diagnostic tests using a variety of high-pass/low-pass filters and concluded that our 1.6icm (48GHz) low-pass filter leaked substantially at the 120 GHz band and above 250 GHz. We suspect that delamination of the filter layers during fabrication may have caused this defect. In the 2022 season, we installed an additional 4icm (120GHz) low pass edge (LPE) filter at the top of the focal plane camera to eliminate this leak(Fig. \ref{fig:fourth}). The out of band dark loading is substantially reduced (by a factor of four) as shown in the histogram in Fig. \ref{fig:fourth}. Additionally, the measured beam profile also shows a significant reduction in the dark pickup (Blue leak) after this upgrade for a dark detector which was common between both deployment seasons (Fig. \ref{fig:fifth}). The beam profile of the 30 GHz detector slightly changes due to the presence of the new filter and the 40 GHz detector beam profile remains the same between both seasons (Fig. \ref{fig:fifth}).  

\FloatBarrier

\subsection{Noise Measurements with On-Sky Loading at the South Pole}
\begin{figure} [ht]
   \begin{center}
   \begin{tabular}{c} 
   \includegraphics[height=7cm]{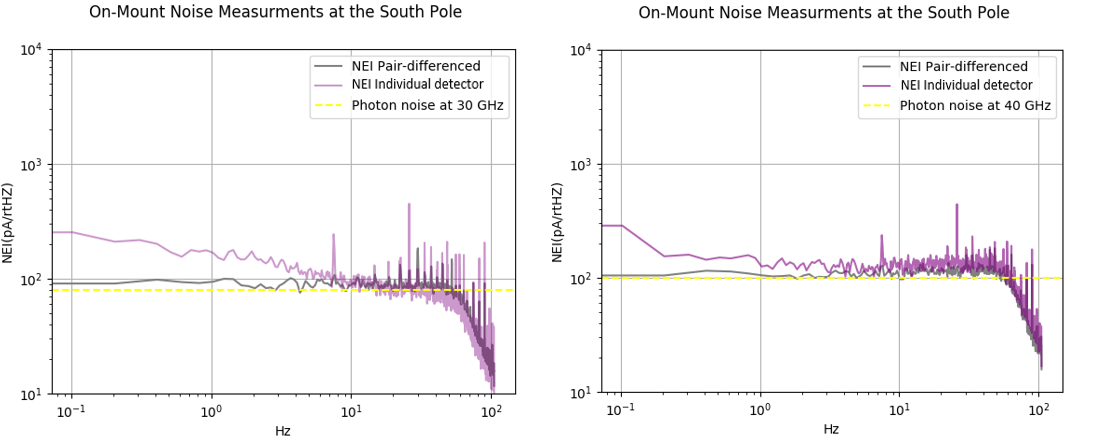}
   \end{tabular}
   \end{center}
   \caption[example] 
   { \label{fig:six} 
   Left: 30 GHz detector. Right: 40 GHz detector. Purple curves show the measured noise equivalent current (NEI) at the South Pole while grey curves show the  difference between pairs in the same pixel. Note that the grey curves agree well with the expected photon noise level shown in yellow, indicating that our detectors are background noise dominated. We biased the detectors in a Titanium superconducting transition.}
   \end{figure}

The BA1 camera sensitivity can be measured using the detector timestreams noise data within the BICEP/Keck science band for sky observation (0.1 Hz to 0.5 Hz). The detectors were designed to be photon noise limited with loading from the South Pole sky. Fig. \ref{fig:six}  shows the 40 GHz and 30 GHz noise spectra before (purple) and after subtracting (grey) detector pairs to eliminate common mode noise. The suppression of 1/f noise down to below 0.1Hz after pair-difference polarization pairs demonstrates that the instability was common mode to both optical detectors. Our experience with other BICEP and Keck cameras leads us to suspect that it likely originates from the atmospheric fluctuations \cite{beam} but the 1/f noise could also arise from other effects. The resulting differenced spectra (grey) agree well with the anticipated photon noise level according to the expected sky loading conditions and calibration parameters for both frequency bands at the South Pole (yellow lines). The resulting spectrum validates photon-noise dominated design. 

\section{CONCLUSIONS} 

The 2022 upgrades to the BA1 camera have improved the performance over that in 2020 deployment season. The measured beam characteristics are in a good agreement with as-designed detector simulations. We also have largely eliminated the high frequency leak in power to boost the overall receiver sensitivity. The on-sky noise measurements show that the instrument is background dominated. These upgrades are important to reach the required sensitivity to map the synchrotron parameters that contributed to the CMB B-modes map. Finally, We plan to deploy the 150 GHz BICEP Array receiver (BA2) this austral summer. We also plan to deploy the 270 GHz BICEP Array receiver as well as the TKIDs demo camera the following year.  

\section{acknowledgments} 
The BICEP/Keck project (including BICEP2, BICEP3, and BICEP Array) have been made possible through a series of grants from the National Science Foundation including 0742818, 0742592, 1044978, 1110087, 1145172, 1145143, 1145248, 1639040, 1638957, 1638978, 1638970, 1726917, 1313010, 1313062, 1313158, 1313287, 0960243, 1836010, 1056465, \& 1255358 and by the Keck Foundation. The development of antenna-coupled detector technology was supported by the JPL Research and Technology Development Fund and NASA Grants 06-ARPA206- 0040, 10-SAT10-0017, 12-SAT12-0031, 14-SAT14-0009, 16-SAT16-0002, \& 18-SAT18-0017. The development and testing of focal planes were supported by the Gordon and Betty Moore Foundation at Caltech. Readout electronics were supported by a Canada Foundation for Innovation grant to UBC. The computations in this paper were run on the Odyssey cluster supported by the FAS Science Division Research Computing Group at Harvard University. The analysis effort at Stanford and SLAC was partially supported by the Department of Energy, Contract DE-AC02-76SF00515. We thank the staff of the U.S. Antarctic Program and in particular the South Pole Station without whose help this research would not have been possible. Tireless administrative support was provided by Kathy Deniston, Sheri Stoll, Irene Coyle, Amy Dierker, Donna Hernandez, and Julie Shih.

\bibliography{report} 

\begin{thebibliography}{10}

\bibitem{sync1}
{Martire1, F. A., Barreiro1. R.B., Martínez-González1, E.},
  ``Characterization of the polarized synchrotron emission from planck and wmap
  data,'' {\em Journal of Cosmology and Astroparticle Physics,}  (2022).

\bibitem{BKI}
{BICEP/Keck Collaboration}, ``{Improved constraints on primordial gravitational
  waves using Planck, WMAP, and BICEP/Keck observations through the 2018
  observing season},'' {\em Phys. Rev. Lett.}  (2021).

\bibitem{delensing}
{BICEP/Keck and SPTpol Collaborations}, ``A demonstration of improved
  constraints on primordial gravitational waves with delensing,'' {\em Phys.
  Rev. D,}  (2020).

\bibitem{hui18}
{Hui, H. et al.}, ``Bicep array: a multi-frequency degree-scale cmb
  polarimeter,'' in [{\em Society of Photo-Optical Instrumentation Engineers
  (SPIE) Conference Series}{\nolinebreak\hspace{0.1em}]},  (2018).

\bibitem{albert}
Wandui, A., e.~a., ``Thermal kinetic inductance detectors for millimeter-wave
  detection,'' {\em Journal of Applied Physics}  (2020).

\bibitem{Ale20}
{Schillaci, A. et al.}, ``Design and performance of the first bicep array
  receiver,'' in [{\em Journal of Low Temperature
  Physics}{\nolinebreak\hspace{0.1em}]},  (2020).

\bibitem{Soliman20}
{Soliman, A. et al.}, ``Design and characterization of antenna-coupled 30/40
  ghz detectors and modules for the bicep array experiment,'' in [{\em Journal
  of Low Temperature Physics}{\nolinebreak\hspace{0.1em}]},  (2020).

\bibitem{Cheng20}
{Zhang, C. et al.}, ``Characterizing the sensitivity of 40 ghz tes bolometers
  for bicep array,'' in [{\em Journal of Low Temperature
  Physics}{\nolinebreak\hspace{0.1em}]},  (2020).

\bibitem{BKII}
{BICEP/Keck Collaboration}, ``{Antenna-coupled TES bolometers used in BICEP2,
  Keck array, and SPIDER},'' {\em The Astrophysical Journal,}  (2015).

\bibitem{soli18}
{Soliman, A. et al.}, ``Design and performance of wide-band corrugated walls
  for the bicep array detector modules at 30/40 ghz,'' in [{\em Society of
  Photo-Optical Instrumentation Engineers (SPIE) Conference
  Series}{\nolinebreak\hspace{0.1em}]},  (2018).

\bibitem{beam1}
{BICEP2 Collaboration, Keck Array Collaboration}, ``Bicep2/keck array xi: Beam
  characterization and temperature-to-polarization leakage in the bk15 data
  set,'' {\em The Astrophysical Journal,}  (2019).

\bibitem{beam}
{BICEP/Keck Collaboration}, ``Bicep/keck xv: The bicep3 cosmic microwave
  background polarimeter and the first three-year data set,'' {\em The
  Astrophysical Journal,}  (2022).

\end{thebibliography}
\bibliographystyle{spiebib} 

\end{document}